\documentclass[useAMS,usenatbib]{mn2e}

\usepackage{graphicx}

\newcommand{\kms}{km s$^{-1}$}

\newcommand{\cmn}{cm$^{-3}$}
\newcommand{\lam}{$\lambda$}
\newcommand{\aj}{AJ} % "Astron. J."
\newcommand{\mnras}{MNRAS} % "Mon. Not. R. Astron. Soc."
\newcommand{\apj}{ApJ} % "Astrophys. J."
\newcommand{\apjl}{ApJ} % "Astrophys. J. Lett."
\newcommand{\apjs}{ApJS} % "Astrophys. J. Suppl. Ser."
\newcommand{\araa}{ARA\&A} % "Ann. Rev. Astron. Astrophys."

\newcommand{\pasp}{PASP} % "Publ. Astron. Soc. Pac."

\newcommand{\nat}{Nat} % ``Nature''
\newcommand{\civ}{\mbox{C\,{\sc iv}}}
\newcommand{\ciii}{\mbox{C\,{\sc iii}}}
\newcommand{\siiv}{\mbox{Si\,{\sc iv}}}
\newcommand{\nv}{\mbox{N\,{\sc v}}}

\title[Emergence of a Quasar Outflow]{Emergence of a Quasar 
Outflow}\author[F. Hamann et al.]{F. Hamann$^{1}$\thanks{E-mail:
hamann@astro.ufl.edu (FH)}, K. F. Kaplan$^{2}$, 
P. Rodr\'iguez Hidalgo$^{1,3}$, J. X. Prochaska$^{2,4}$ \newauthor 
and S. Herbert-Fort$^{5}$\\
$^{1}$Department of Astronomy, University of Florida, Gainesville, FL 
32611-2055, USA\\
$^{2}$Department of Astronomy and Astrophysics,
University of California, Santa Cruz, Santa Cruz, CA 95064, USA\\
$^{3}$Department of Astronomy and Astrophysics, Pennsylvania State University,
525 Davey Lab, University Park, PA 16802, USA\\
$^{4}$University of California Observatories -
Lick Observatory, University of California, Santa Cruz, CA 95064, USA\\
$^{5}$University of Arizona/Steward Observatory, 933 N Cherry Avenue,
Tucson, AZ 85721, USA}

\begin{document}

\date{Accepted 2008 August 22. Received 2008 August 20; 
in original form 2008 August 14}

\pagerange{\pageref{firstpage}--\pageref{lastpage}} \pubyear{2008}

\maketitle

\label{firstpage}

\begin{abstract}
We report the first discovery of the emergence of a high-velocity 
broad-line outflow in a luminous quasar, J105400.40+034801.2 at
redshift $z\sim 2.1$. The outflow is evident in ultraviolet 
\civ\ and \siiv\ absorption lines with velocity shifts $v\sim 26,300$ 
\kms\ and deblended widths FWHM $\sim$ 4000 \kms . These features 
are marginally strong and broad enough to be considered broad 
absorption lines (BALs), but their large velocities exclude them from 
the standard BAL definition. The outflow lines appeared between two 
observations in the years 2002.18 and 2006.96. A third observation 
in 2008.48 showed the lines becoming $\sim$40\% weaker 
and 10\% to 15\% narrower. There is no evidence for acceleration 
or for any outflow gas at velocities $\la$23,000 \kms . 
The lines appear to be optically thick, with 
the absorber covering just 20\% of the quasar continuum 
source. This indicates a characteristic absorber size of 
$\sim$$4\times 10^{15}$ cm, but with a BAL-like total column density 
$\log N_H ({\rm cm}^{-2})\ga 21.2$ and average space density 
$n_H\ga 2\times 10^5$ \cmn . We attribute the emergence 
of the outflow lines to a substantial flow structure 
moving across our line of sight, possibly near the ragged edge 
of the main BAL flow or possibly related to the onset of 
a BAL evolutionary phase.
\end{abstract}


\section{Introduction}

Accretion disk outflows are an important part of the quasar 
phenomenon. They are known to be 
present in at least 50\% of optically-selected quasars  
\citep{Hamann04,Nestor08,Ganguly08,Paola08}. 
They might play a critical role in the accretion 
physics and in the observational properties of quasars, and 
they might expel enough metal-rich gas and kinetic energy 
to have a major impact on star formation and galaxy evolution 
in their surroundings \citep{Kauffmann00,Richards02,
diMatteo05,Everett05,Proga05}. The outflows appear most 
conspicuously in quasar spectra as 
broad absorption lines (BALs), which typically have velocity 
widths of 5000--20,000 \kms . However, there is 
also a variety of weaker and narrower outflow lines that are just 
as common as BALs. These include the so-called ``mini''-BALs, which 
appear at velocities from near 0 to $\sim$0.2$c$ with profile 
widths from the BAL range down to a few hundred \kms . 
Understanding the relationships between the various outflow 
lines is fundamental to our understanding of the outflows themselves. 
For example, in one scenario, the different line types represent 
different manifestations of a single outflow phenomenon viewed 
at different angles \citep{Elvis00,Ganguly01}. Alternatively, 
they might represent an evolutionary 
sequence, where weak mini-BALs appear near the beginning or 
end of a more powerful BAL outflow phase \citep{Hamann04}.

Variability studies can help test these ideas by providing 
constraints on the outflow dynamics, stability, location and 
basic physical properties. Recent studies have shown, for example, 
that mini-BALs tend to be more variable than the 
broader and stronger BALs 
\citep{Gibson08,Paola08,Capellupo08}. Approximately half of 
mini-BALs show significant variations on time scales from 
a few years to few months in the quasar rest-frame. In rare 
cases, weak mini-BALs become much stronger and broader 
like BALs, or they disappear altogether. 
Following the emergence of a quasar outflow would be 
particularly valuable, but we know almost nothing 
about such occurances because variability programs start with 
quasars that are already known to have outflow lines. The only 
known case of an emerging BAL outflow was in a 
narrow line Seyfert 1 galaxy at redshift $\sim$0.03  
\citep{Leighly05,Leighly08}. 
The dramatic appearance of BALs in this object is 
surprising because it is much less luminous 
(by a factor of $\ga$100) than typical BAL quasars.  

In this paper, we report the first discovery of an 
emerging high-velocity BAL-like outflow in a luminous quasar, 
J105400.40+034801.2 at redshift $\sim$2.1. We observed this 
quasar as part of a program to study metal-strong 
damped Ly$\alpha$ (DLA) absorption systems 
\citep{Herbert-Fort06,Kaplan08}. Those data showed the new 
appearance of broad outflow lines compared to spectra 
obtained several years earlier by the Sloan Digital 
Sky Survey (SDSS). 
In the sections below we discuss the observations (\S2), the 
measured properties of the outflow lines (\S3), and some 
theoretical implications of our results (\S4). 
 
\section[]{Observations}

The Sloan Digitial Sky Survey (SDSS) obtained the first high-quality 
spectrum of J105400.40+034801.2 on 5 March 2002 (2002.18). 
We retrieved the fully reduced spectrum from the SDSS archives,  
Data Release 6. This spectrum covers 
wavelengths from 3805 \AA\ to 9210 \AA\ at resolution $R\equiv 
\lambda/\Delta\lambda\approx 2000$ (150 \kms ). The 
absolute fluxes measured through through a $\sim$3$^{\prime\prime}$ 
aperture fiber are expected to be accurate to within a few percent 
\citep{Adelman08}. The emission line redshift and the blue and red 
magnitudes reported by the SDSS are $z_{em}=2.095$, $g=18.11$ and 
$r=17.98$, respectively. We also note that this quasar is radio-quiet 
based on a non-detection in the FIRST radio survey \citep{Becker95}.
 
We observed J105400.40+034801.2 on 16 December 2006 (2006.96) 
using the Multi-Mirrored Telescope (MMT) Spectrograph in the blue 
channel with the 800 groove/mm grating. Combined with a 
1.5$^{\prime\prime}$ wide slit, this setup provided 
wavelength coverage from 3090 \AA\ to 5095 \AA\ at resolution 
$R \approx 1370$ (220 \kms ). The exposure time was 720 s. 
After noticing the broad outflow lines in the MMT data, 
we obtained another spectrum on 30 May 2008 (2008.48) 
using the Low Resolution Imaging Spectrograph (LRIS) at the Keck 
Observatory. In this case, a 600 groove/mm grating blazed at 6000 
\AA\ and a 1$^{\prime\prime}$ entrance slit provided wavelength 
coverage from 3105 \AA\ to 5600 \AA\ at resolution $R\approx 1000$ 
(300 \kms ). The total exposure time was 400 s. The observations 
at the MMT and Keck were both performed 
with the slit at the parallactic angle. The spectra were extracted 
and processed by standard techniques using the Low Redux 
Pipeline\footnote{http://www.ucolick.org/$\sim$xavier/LowRedux/index.html} 
software package. The relative fluxes were calibrated using 
spectrophotometric standards observed on the same night. 
The absolute fluxes have uncertainties up to 50\%. 
 
\section[]{Results}
 
Figure 1 shows the three spectra obtained in 2002.18 (SDSS), 
2006.96 (MMT) and 2008.41 (Keck). The wavelengths in the 
quasar rest frame are based on $z_{em}=2.095$ reported by the 
SDSS. The vertical flux scale in the figure  
applies to the SDSS spectrum. The other spectra are scaled  
vertically to match the SDSS fluxes at continuum wavelengths. 
The SDSS and MMT spectra are shown after smoothing for 
easier display. This smoothing has no effect on the appearance 
of broad features like the outflow lines and the broad emission 
lines (BELs).
 
 \begin{figure*}
 \includegraphics[scale=0.65]{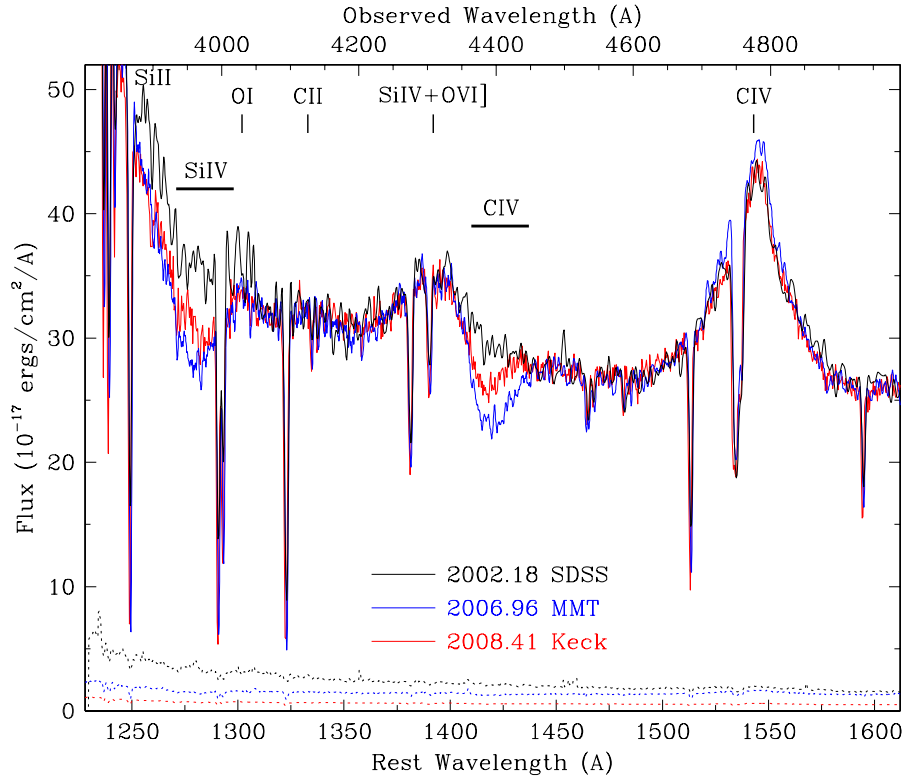}
% \vspace{3pt}
 \caption{Spectra of J105400.40+034801.2  at rest-frame wavelengths 
(bottom scale) and observed (top) are represented by the black, 
blue and red curves for the three observations indicated. 
The broad \siiv\ and \civ\ outflow lines are marked by horizontal 
bars above the data. The locations of BELs are marked across 
the top. The steep rise at short wavelengths is due 
to strong \nv\ and Ly$\alpha$ emission. 
All of the strong narrow absorption lines (not labeled) belong 
to an unrelated DLA system at $z_{abs}=2.0683$. 
The vertical flux scale applies to the 
SDSS spectrum. The other spectra are scaled vertically 
 to match. The formal 1$\sigma$ errors are shown 
 by dotted curves near the bottom.}
\end{figure*}

Broad high-velocity outflow lines are clearly present in 
\siiv\ \lam\lam 1394,1403 
and \civ\ \lam\lam 1548,1551 in the 2006.96 and 2008.48 spectra. 
They appeared in $\leq$1.54 yr (rest frame) 
between the 2002.18 and 2006.96 observations, and then they weakened 
again 0.47 years later in the 2008.48 measurement. Meanwhile, the 
BEL profiles and equivalent widths (including Ly$\alpha$ and 
\ciii ] \lam 1909 not shown in Fig. 1) 
varied by $\la$5\%. The discovery spectrum of J105400.40+034801.2 
obtained in 1988.15 \citep{Clowes94} is much noisier and 
has a lower resolution ($R\sim 440$) than the data shown in Figure 1. 
Outflow lines like the ones measured in 2006.96 would probably 
not be detectable in those data. However, the 1988.15 spectrum does 
rule out the presence of broad absorption features with $\sim$2 or 
more times the strength of the lines in 2006.96. No other 
outflow lines in the broad outflow system are detected in any of 
the data.  The 2006.96 and 2008.48 spectra both have a 
depression at the position of \nv\ \lam\lam 1239,1243 
that is consistent with the broad absorption in \civ\ and \siiv . 
However, we cannot confirm the reality of 
this feature because the spectra at those wavelengths are severely 
contaminated by Ly$\alpha$ forest lines. 
  
To measure the outflow 
lines quantitatively, we define a ``pseudo"-continuum 
across the true quasar continuum and the tops of the BELs. This 
pseudo-continuum matches the 2008.48 spectrum except at the 
wavelengths of the broad absorption, where it follows the 
2002.18 data.  
The result is shown in the top panel of Figure 2. Notice that
the pseudo-continuum ignores the bump in the 2002.18 SDSS spectrum 
around 3900 \AA\ (1260 \AA\ rest). We attribute this bump  
to an anomaly in the SDSS flux calibration. Our experience with 
many similar SDSS spectra 
indicates that anomalies like this do sometimes occur near 
the blue edge of the SDSS wavelength coverage. In any case, this 
feature is not related to the \siiv\ outflow line because it has no 
analog near the \civ\ feature. 

\begin{figure}
 \includegraphics[scale=0.49]{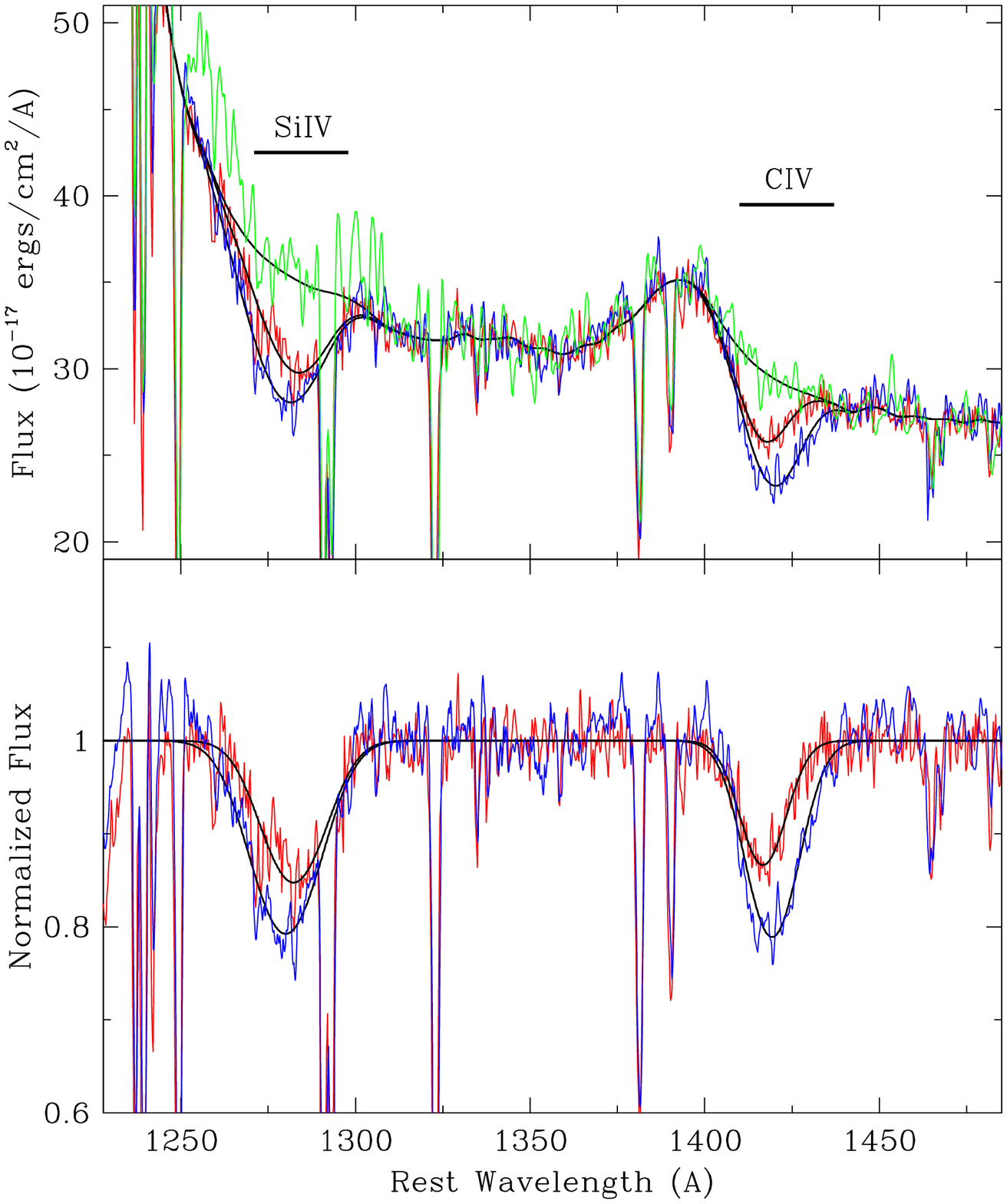}
% \vspace{302pt}
 \caption{Top panel: Spectra of J105400.40+034801.2 in the same 
format as Figure 1, but with the SDSS data now drawn as a 
green curve and with fits to the pseudo-continuum and the 
outflow absorption lines shown by the smooth 
 black curves. Bottom panel: Normalized spectra from 2006.96 (blue) 
and 2008.48 (red) together with the normalized line fits (black).}
\end{figure}

The bottom panel in Figure 2 shows the 2006.96 and 2008.48  
spectra normalized by the pseudo-continuum. We fit the broad 
\siiv\ and \civ\ absorption lines in these normalized spectra using 
gaussian optical depth profiles. The \siiv\ and \civ\ transitions are 
doublets with velocity separations of $\sim$500 \kms\ and 
$\sim$1930 \kms , respectively. 
Our fits include one gaussian per doublet member, with the redshifts 
and doppler velocities tied together. We assume that the 
doublet members have equal strengths based on the evidence  
for line saturation described in \S4.1. In our fitting procedure, 
this means forcing the optical depths within the doublets to be equal   
even though their actual ratios based on oscillator strengths 
should be 2:1. The fits assume implicitly that the absorbing gas 
fully covers the quasar emission regions. We will discuss the effects 
of partial covering and line saturation in \S4.1 below. 

Figure 2 shows the final fits to the broad outflow lines. 
Table 1 lists some of the derived results, including 
the rest equivalent width (REW) of the entire blended doublet, 
and the absorption redshift ($z_{abs}$), the velocity 
shift ($v$ measured from $z_{em} = 2.095$), the full width 
at half minimum (FWHM) of the individual doublet lines, and  
the logarithmic ionic column 
density ($\log N$) derived from the weaker doublet members. 
Note that the column densities are lower limits because 
of probable saturation effects (\S4.1).  

\begin{table}
 \centering
 \begin{minipage}{77mm}
  \caption{Outflow Line Measurements}
  \begin{tabular}{@{}lccccc@{}}
  \hline
   Line & REW & $z_{abs}$ & $v$ & FHWM 
     & $\log N$ \\
    & (\AA) &  & (\kms ) & (\kms ) 
     & (cm$^{-2}$) \\
\hline
   \multicolumn{6}{c}{------------------------ 2006.96 (MMT) ------------------------}\\
 Si IV & 6.21 & 1.833 & 26,430 & 4830 & $\geq$14.84 \\
 C IV & 4.83 & 1.835 & 26,270 & 3660 & $\geq$15.08 \\
% \vspace{5pt}
 
    \multicolumn{6}{c}{------------------------ 2008.48 (Keck) ------------------------}\\
 Si IV & 4.07 & 1.838 & 25,900 & 4330 & $\geq$14.66 \\
 C IV & 2.54 & 1.830 & 26,810 & 3160 & $\geq$14.80\\
\hline
\end{tabular}
\end{minipage}
\end{table}

The broad outflow lines in 2006.96 had velocity 
shifts $v\sim 26,300$ \kms\ and deblended FWHMs near 
$\sim$4000 \kms . The actual measured width of the \civ\ blend  
was FWHM $\sim$ 4320 \kms . These features are strong and broad 
enough to be considered BALs. However, their large velocity shifts 
lead to a balnicity index of zero, which excludes 
them from the standard BAL definition 
\citep{Weymann91}. Their appearance 
in 2006.96 corresponds to strengthening by at least a factor 
of $\sim$5 compared to the 2002.18 observation. 
Then they became roughly $\sim$40\% weaker and 10\% to 15\% 
narrower by 2008.48. 

The uncertainties in our measured results are difficult to assess. 
The \civ\ data should be more reliable than \siiv\ because 
the spectra at the \civ\ wavelengths are less noisy and there is 
less blending with underlying BELs. There is no obvious explanation 
for the different FWHMs and velocity shifts reported for \civ\ and 
\siiv\ lines in Table 1 in terms of our fits to the lines or 
the pseudo-continuum. These differences might be real, or they might 
be indications of the true measurement uncertainties, mostly in 
the \siiv\ line. 

\section{Discussion}

\subsection{Covering Fractions and Optical Depths}

If the outflow gas does not fully cover the background 
emission regions, then our line fits in \S3 will underestimate 
the true optical depths and column densities. Partial 
covering appears to be common in BALs and it is well documented 
in many narrow-line outflow systems where resolved doublets 
provide direct measures of the covering fractions and 
true optical depths \citep{Hamann97,Hamann98,Ganguly99,
Arav01,Hamann04,Gabel05}. The net result can be 
shallow optically thick lines whose strengths and profiles 
are controlled by a velocity-dependent covering fraction. 

In the J105400.40+034801.2 outflow, we cannot constrain the 
true optical depths based on resolved doublet ratios. 
However, standard photoionization models applied to quasar outflows 
predict that the optical depth in \civ\ should be at least 5 times 
larger than \siiv\ if the Si/C abundance is roughly solar and the degree 
of ionization favors Si$^{+3}$ and C$^{+3}$ over  
other ions \citep{Hamann97d,Hamann99}. The observed \siiv /\civ\ 
absorption ratio 
near unity in weak lines strongly suggests that both features are 
saturated. Their measured line depths then imply that the covering 
fraction was $\sim$20\% (at line center) in 2006.96. In addition, 
the velocity shift of the outflow lines leads to little or no 
absorption at the BEL wavelengths (Figure 1). Therefore, the 
$\sim$20\% covering fraction applies to the background continuum 
source. 

\subsection{Physical Properties and Time Scales}

We can place some interesting constraints on the outflow physical 
properties by making a few reasonable assumptions. For example, 
in the photoionization calculations just mentioned,  
our lower limit on the Si$^{+3}$ column density 
in 2006.96 (Table 1) corresponds to a total hydrogen column of 
$\log N_H ({\rm cm}^{-2})\ga 19.9$ if the Si/H abundance is 
roughly solar. In the more likely event that 
the \siiv\ line is saturated with line-center optical 
depth $\tau_o \ga 3$ (\S4.1), the total column density in the 
absorber would be $\log N_H ({\rm cm}^{-2})\ga 21.2$. 

Partial covering of the quasar continuum source (\S4.1) implies 
that the absorbing region is small. The radius    
of the accretion disk continuum source at 1550 \AA\ is expected to 
be roughly $R_{1550}\sim 10^{16}$ cm in luminous quasars 
like J105400.40+034801.2\footnote{We 
estimate the bolometric luminosity of J105400.40+034801.2 
to be $L\sim 10^{47}$ ergs s$^{-1}$ based on the observed SDSS 
flux at 1450 \AA\ (rest), a cosmology with $H_o = 71$ \kms\ Mpc, 
$\Omega_M = 0.27$, $\Omega_{\Lambda}=0.73$ and a standard 
bolometric correction factor $L\approx 3.4\lambda 
L_{\lambda}(1450 {\rm \AA})$. This implies 
characteristic radii for the continuum source 
at 1550 \AA\ of $R_{1550}\sim 10^{16}$ cm and for 
the \civ\ BEL region of $R_{\rm CIV}\sim 6\times 10^{17}$ cm
\citep{Peterson04,Bentz07,Gaskell08,Hamann08}.}. 
In simple geometries, the 20\% 
covering fraction indicates a characteristic absorber radius of  
$\sim$$4\times 10^{15}$ cm. (We do not consider the possibility 
of continuum flux scattered from some larger region because 
the $\sim$20\% covering fraction would require that $\ga$80\% 
of the observed continuum is scattered light.) 
Comparing this size to the 
minimum column density above indicates further that the average 
volume density is $n_H\ga 2\times 10^5$ \cmn . 

The outflow dynamical times are interesting for comparison 
to the observed variabilities. BAL-like outflows are 
believed to be launched from a rotating accretion disk, and 
the \civ\ absorption lines might form at radii just beyond the  
\civ\ BEL region \citep{Murray97,Proga00,Everett05}. A characteristic 
flow time in this situation would be $t_f\sim R_{\rm CIV}/v\sim 7$ yr, 
where $v\sim 26,300$ \kms\ is the flow speed and $R_{\rm CIV}$ 
is the CIV BEL region radius$^2$. The transit time 
for absorbing clouds crossing the 1550 \AA\ continuum 
source depends on the transverse velocities in the flow (i.e., 
perpendicular to our lines of sight). If the transverse velocities  
in the \siiv\ and \civ\ absorbing gas are similar to the 
disk rotation speed just beyond $R_{\rm CIV}$, roughly 
$v_{tr}\sim 3000$ \kms , then the transit time would be 
$t_{tr}\sim R_{1550}/v_{tr} \sim 1$ yr. Thus it appears that 
the transit time is more compatible with the observations 
than the overall flow time.

\subsection{What Caused the Outflow Line Variations?}

Did a new BAL-like outflow emerge from a situation where 
there was previously no outflow at all? We cannot answer that  
question except to note that there were no indications 
of structural changes in the 
accretion disk that might signal the onset of an outflow. 
In particular, 
the BELs and near-UV spectral slope did not change significantly 
during the absorption line variability period, and we can 
rule out large (factor of $\ga$2) changes in the near-UV flux. 
There are also no clear signs of acceleration. 
If a new outflow did emerge in J105400.40+034801.2, it probably 
appeared in our line of sight as Si$^{+3}$ 
and C$^{+3}$ after the acceleration to $v\sim 26,300$ \kms\ 
had already occurred. 

The emergence of outflow lines might have been caused 
by changes in the 
ionization or by the movement of gas across our lines 
of sight. Changes in the ionization are an unlikely 
explanation for several reasons. First, the \civ\ and \siiv\ BELs 
did not vary. Their ionization is controlled by the same far-UV 
flux that controls the \civ\ and \siiv\ absorption lines. 
Changes in quasar 
continuum fluxes are known to cause changes in the BELs after 
a time lag related to the light travel time between the continuum 
source and the BEL region. The lag time for the \civ\ BEL in 
J105400.40+034801.2 should be about half a year 
in the rest frame$^2$. Therefore, substantial changes in the 
far-UV flux should have been evident in the CIV BEL. Second, weak 
\siiv\ and \civ\ absorption lines affected by 
changes in the ionization should exhibit measurable changes in the 
\siiv /\civ\ line ratio (see the photoionization calculations 
mentioned in \S4.1). This did not occur in J105400.40+034801.2 
because the lines are optically thick. Their strengths depend on 
the covering fraction much more than the degree of 
ionization. 

We conclude that the outflow line variations were  
caused by changes in the covering fraction related to the 
movement of gas across our line of sight (see also 
\citealt{Gibson08}). Our crude estimate 
of the crossing time, $\sim$1 yr (\S4.2), is broadly consistent 
with this conclusion. Instabilities might also 
play a role by producing transient flow structures 
capable of \civ\ and \siiv\ absorption. 
However, instabilities might be expected 
evolve on roughly a flow time, $\sim$7 yr (\S4.2), which is 
perhaps too slow to explain the observations. 
Better assessments of the origins of the variability will 
require detailed hydrodynamic simulations.  

\section{Summary and Conclusions}

We have described the first known case of an emerging broad-line  
outflow in a luminous high-redshift quasar, J105400.40+034801.2. 
The outflow is evident in \siiv\ and \civ\ absorption lines having 
FWHM $\sim 4000$ \kms\ and velocity shifts $v\sim 26,300$ \kms . 
These features are not technically BALs because their large 
velocity shifts lead to a balnicity index of zero. 
Nonetheless, they represent a powerful 
outflow with $\log N_H ({\rm cm}^{-2})\ga 21.2$ and  
$n_H\ga 2\times 10^5$ \cmn . The outflow lines appeared 
between two observations 1.54 yr apart 
in the quasar rest frame. A third observation 0.47 yr later shows 
the lines becoming weaker again by $\sim$40\%. We attribute this 
behavior to a substantial flow structure crossing our line of 
sight to the continuum source. 

Overall, the outflow lines discussed here are 
similar to high-velocity mini-BALs observed in other quasars 
\citep{Paola08}. These features are characteristically 
narrower, weaker and more variable than the classic strong BALs. 
One possible explanation for these outflow lines is that they 
form along the ragged edge of 
the main BAL outflow, where the flow structures could 
be naturally smaller and more volatile. If BAL outflows are 
typically clumpy or filamentary \citep{Hall07}, then perhaps 
the mini-BALs represent individual clumps or filaments 
along the edges of these flows. 
Another possibility is that mini-BALs and other similar 
outflow lines represent the sputtering 
beginning or end stages of a BAL evolutionary phase. 
Continued monitoring of quasars like J105400.40+034801.2, e.g., 
with multi-wavelength measurements to get better constraints 
on the ionization and column densities, will help to test 
these theoretical ideas.

\section*{Acknowledgments}

We are grateful to Karen Leighly, Daniel Capellupo, Doron 
Chelouche, Daniel Proga and an anonymous referee for helpful 
comments. 

\bibliographystyle{mn2e}
%\bibliography{bibliography}

\begin{thebibliography}{}

\bibitem[\protect\citeauthoryear{Adelman-McCarthy et 
al.}{2008}]{Adelman08} Adelman-McCarthy J.~K., et al., 2008, 
ApJS, 175, 297 

\bibitem[\protect\citeauthoryear{Arav et al.}{2001}]{Arav01} 
Arav N., et al., 2001, ApJ, 561, 118 

\bibitem[\protect\citeauthoryear{Becker, White, 
\& Helfand}{1995}]{Becker95} Becker R.~H., White R.~L., Helfand D.~J., 1995, ApJ, 450, 559 

\bibitem[\protect\citeauthoryear{Bentz et al.}{2007}]{Bentz07} 
Bentz M.~C., Denney K.~D., Peterson B.~M., Pogge R.~W., 2007, ASPC, 373, 
380 

\bibitem[\protect\citeauthoryear{Capellupo et~al.}{2008}]{Capellupo08}
{Capellupo} D.,  {Hamann} F., {Rodr\'iguez Hidalgo} P.,    {Shields} J.,  2008,
  in prep.

\bibitem[\protect\citeauthoryear{{Clowes} \& {Campusano}}{{Clowes} \&
  {Campusano}}{1994}]{Clowes94}
{Clowes} R.~G.,  {Campusano} L.~E.,  1994, \mnras, 266, 317

\bibitem[\protect\citeauthoryear{{Di Matteo}, {Springel} \& {Hernquist}}{{Di
  Matteo} et~al.}{2005}]{diMatteo05}
{Di Matteo} T.,  {Springel} V.,    {Hernquist} L.,  2005, \nat, 433, 604

\bibitem[\protect\citeauthoryear{{Elvis}}{{Elvis}}{2000}]{Elvis00}
{Elvis} M.,  2000, \apj, 545, 63

\bibitem[\protect\citeauthoryear{{Everett}}{{Everett}}{2005}]{Everett05}
{Everett} J.~E.,  2005, \apj, 631, 689

\bibitem[\protect\citeauthoryear{Gabel et al.}{2005}]{Gabel05} 
Gabel J.~R., et al., 2005, ApJ, 623, 85 

\bibitem[\protect\citeauthoryear{Ganguly et 
al.}{1999}]{Ganguly99} Ganguly R., Eracleous M., Charlton J.~C., 
Churchill C.~W., 1999, AJ, 117, 2594 

\bibitem[\protect\citeauthoryear{{Ganguly}, {Bond}, {Charlton}, {Eracleous},
  {Brandt} \& {Churchill}}{{Ganguly} et~al.}{2001}]{Ganguly01}
{Ganguly} R.,  {Bond} N.~A.,  {Charlton} J.~C.,  {Eracleous} M.,  {Brandt}
  W.~N.,    {Churchill} C.~W.,  2001, \apj, 549, 133

\bibitem[\protect\citeauthoryear{{Ganguly} \& {Brotherton}}{{Ganguly} \&
  {Brotherton}}{2008}]{Ganguly08}
{Ganguly} R.,  {Brotherton} M.~S.,  2008, \apj, 672, 102

\bibitem[\protect\citeauthoryear{Gaskell}{2008}]{Gaskell08} 
Gaskell C.~M., 2008, RMxAC, 32, 1 

\bibitem[\protect\citeauthoryear{{Gibson}, {Brandt}, {Schneider} \&
  {Gallagher}}{{Gibson} et~al.}{2008}]{Gibson08}
{Gibson} R.~R.,  {Brandt} W.~N.,  {Schneider} D.~P.,    {Gallagher} S.~C.,
  2008, \apj, 675, 985

\bibitem[\protect\citeauthoryear{{Hall}, {Sadavoy}, {Hutsemekers}, {Everett} \&
  {Rafiee}}{{Hall} et~al.}{2007}]{Hall07}
{Hall} P.~B.,  {Sadavoy} S.~I.,  {Hutsemekers} D.,  {Everett} J.~E.,
  {Rafiee} A.,  2007, \apj, 665, 174

\bibitem[\protect\citeauthoryear{{Hamann}}{{Hamann}}{1997}]{Hamann97d}
{Hamann} F.,  1997, \apjs, 109, 279

\bibitem[\protect\citeauthoryear{{Hamann}}{{Hamann}}{1998}]{Hamann98}
{Hamann} F.,  1998, \apj, 500, 798

\bibitem[\protect\citeauthoryear{{Hamann}\& {Simon}}{{Hamann} \& 
{Simon}}{2008}]{Hamann08}{Hamann} F., {Simon} L. E., 2008, in prep.

\bibitem[\protect\citeauthoryear{{Hamann}, {Barlow}, {Junkkarinen} \&
  {Burbidge}}{{Hamann} et~al.}{1997}]{Hamann97}
{Hamann} F.,  {Barlow} T.~A.,  {Junkkarinen} V.,    {Burbidge} E.~M.,  1997,
  \apj, 478, 80

\bibitem[\protect\citeauthoryear{{Hamann} \& {Ferland}}{{Hamann} \&
  {Ferland}}{1999}]{Hamann99}
{Hamann} F.,  {Ferland} G.,  1999, \araa, 37, 487

\bibitem[\protect\citeauthoryear{Hamann 
\& Sabra}{2004}]{Hamann04} Hamann F., Sabra B., 2004, ASPC, 311, 203 

\bibitem[\protect\citeauthoryear{{Herbert-Fort}, {Prochaska},
  {Dessauges-Zavadsky}, {Ellison}, {Howk}, {Wolfe} \&
  {Prochter}}{{Herbert-Fort} et~al.}{2006}]{Herbert-Fort06}
{Herbert-Fort} S.,  {Prochaska} J.~X.,  {Dessauges-Zavadsky} M.,  {Ellison}
  S.~L.,  {Howk} J.~C.,  {Wolfe} A.~M.,    {Prochter} G.~E.,  2006, \pasp, 118,
  1077

\bibitem[\protect\citeauthoryear{{Kaplan} \& {et al.}}{{Kaplan} \& {et
  al.}}{2008}]{Kaplan08}
{Kaplan} K.~F.,  {et al.} 2008, in prep.

\bibitem[\protect\citeauthoryear{{Kauffmann} \& {Haehnelt}}{{Kauffmann} \&
  {Haehnelt}}{2000}]{Kauffmann00}
{Kauffmann} G.,  {Haehnelt} M.,  2000, \mnras, 311, 576

\bibitem[\protect\citeauthoryear{{Leighly}}{{Leighly}}{2008}]{Leighly08}
{Leighly} K.,  2008, \apj, (submitted)

\bibitem[Leighly et al.(2005)]{Leighly05} Leighly, K.~M., Casebeer, D.~A., Hamann, F., \& Grupe, D.\ 2005, Bulletin of the American Astronomical Society, 37, 1184

\bibitem[\protect\citeauthoryear{{Murray} \& {Chiang}}{{Murray} \&
  {Chiang}}{1997}]{Murray97}
{Murray} N.,  {Chiang} J.,  1997, \apj, 474, 91

\bibitem[\protect\citeauthoryear{{Nestor}, {Hamann} \& {Rodriguez
  Hidalgo}}{{Nestor} et~al.}{2008}]{Nestor08}
{Nestor} D.,  {Hamann} F.,    {Rodriguez Hidalgo} P.,  2008, \mnras, 386, 2055

\bibitem[\protect\citeauthoryear{Peterson et 
al.}{2004}]{Peterson04} Peterson B.~M., et al., 2004, ApJ, 613, 
682 

\bibitem[\protect\citeauthoryear{Proga, Stone, 
\& Kallman}{2000}]{Proga00} Proga D., Stone J.~M., Kallman T.~R., 2000, 
ApJ, 543, 686 

\bibitem[\protect\citeauthoryear{{Proga}}{{Proga}}{2005}]{Proga05}
{Proga} D.,  2005, \apjl, 630, L9

\bibitem[\protect\citeauthoryear{{Richards}, {Vanden Berk}, {Reichard}, {Hall},
  {Schneider}, {SubbaRao}, {Thakar} \& {York}}{{Richards}
  et~al.}{2002}]{Richards02}
{Richards} G.~T.,  {Vanden Berk} D.~E.,  {Reichard} T.~A.,  {Hall} P.~B.,
  {Schneider} D.~P.,  {SubbaRao} M.,  {Thakar} A.~R.,    {York} D.~G.,  2002,
  \aj, 124, 1

\bibitem[\protect\citeauthoryear{{Rodr\'iguez Hidalgo}, {Hamann}, {Nestor} \&
  {Shields}}{{Rodr\'iguez Hidalgo} et~al.}{2008}]{Paola08}
{Rodr\'iguez Hidalgo} P.,  {Hamann} F.,  {Nestor} D.~B.,    {Shields} J.,
  2008, in prep.

\bibitem[\protect\citeauthoryear{{Weymann}, {Morris}, {Foltz} \&
  {Hewett}}{{Weymann} et~al.}{1991}]{Weymann91}
{Weymann} R.~J.,  {Morris} S.~L.,  {Foltz} C.~B.,    {Hewett} P.~C.,  1991,
  \apj, 373, 23

\end{thebibliography}

\bsp

\label{lastpage}

\end{document}